\documentclass[pre,preprint, groupedaddress,showkeys, showpacs]{revtex4}

\usepackage{graphicx}
\begin{document}


\title{Competition of spatial and temporal instabilities under time delay near codimension-two Turing-Hopf bifurcations}


\author{Huijuan Wang$^{ }$}\email[Email:]{whuijuan@yahoo.com.cn}
\author{Zhi Ren}


\affiliation{School of Mathematics and Physics, North China Electric Power University, Baoding 071003, China}


\date{\today}
\begin{abstract}
\indent Competition of spatial and temporal instabilities under time delay near the codimension-two Turing-Hopf bifurcations is studied in a reaction-diffusion equation. The time delay changes remarkably the oscillation frequency, the intrinsic wave vector, and the intensities of both Turing and Hopf modes. The application of appropriate time delay can control the competition between the Turing and Hopf modes. Analysis shows that individual or both feedbacks can realize the control of the transformation between the Turing and Hopf patterns. Two dimensional numerical simulations validate the analytical results.
\end{abstract}


\maketitle

\section{Introduction}
\indent Spatiotemporal patterns have been of great interest in a variety of biological, chemical, and physical contexts \cite{Cross}. From the viewpoint of dynamics of pattern formation, these patterns can be classified as three types according to the instability mechanisms: i) periodic in space but stationary in time, ii) oscillatory in time but uniform in space (in some cases, which often gives rise to phase waves due to the couple of the phases of the oscillators), iii) periodic in space and oscillatory in time. Based on linear stability analysis of a homogeneous state, these instabilities correspond to Turing, Hopf, and wave bifurcation, respectively. Recently, the interactions between these instabilities, which often give rise to novel phenomena of pattern formation, have attracted much attention \cite{Yang1, Yang2, Ricard, DeWit, Yang3, Dilao, Meixner, Just, Bose, Rovinsky, Yuan}. Yang $et$ $al$ studied the interactions between the Turing and wave instabilities and observed the coexistence of Turing spots and antispirals \cite{Yang1}. Interaction of two Turing modes leads to the formation of "black-eyes" pattern and the superposition patterns combining stripes and/or spots \cite{Yang2}. Of particular interests are the interaction between the periodic Turing mode and the oscillatory Hopf mode \cite{Ricard}. It has been explored in a Brusselator model\cite{DeWit, Yang3, Dilao}, in a semiconductor model \cite{Meixner, Just, Bose}, and in a Lengyel-Epstein model \cite{Rovinsky, Yuan}. The competition of the two modes near codimension-two Turing-Hopf bifurcations gives rise to the oscillatory Turing patterns, the stable squares, and the bistability between homogeneous oscillatory states and hexagonal Turing patterns.

\indent Time-delayed-feedback is a practical method for controlling the dynamics of a system \cite{Murray,Mikhailov}. It is based on applying feedback perturbation proportional to the deviation of the current state from the delayed state of the system. It was originally presented by Ott $et$ $al$ to control the chaotic behavior of a deterministic system \cite{Ott}. Recently, time-delayed-feedback has been used to control the pattern formations in spatially extended systems. The time delay with different feedback forms affects the behaviors of pattern formations much \cite{Zykov2, Naknaimueang, Kheowan, Li1, Li2, Li3, Veflingstad, Pollmann,Guo}. It can stabilize the rigid rotation of a spiral or completely destroy a spiral \cite{Zykov2}. The tip trajectories of spirals can be controlled by applying time-delayed-feedback \cite{Naknaimueang, Kheowan}. The spontaneous suppression of spiral turbulence by using feedback has been studied experimentally in a light-sensitive Belousov-Zhabotinsky reaction and numerically in a modified FitzHugh-Nagumo model \cite{Guo}. Based on a Brusselator reaction-diffusion model, Li $et$ $al$ studied the effect of time delay on various pattern formations, such as the traveling and standing patterns, the inward and outward spiral waves \cite{Li1, Li2, Li3}.

\indent Just recently, Sen $et$ $al$ studied the transition between the Turing and Hopf instabilities in a reaction-diffusion model, in which the time delay is incorporated in the kinetic terms \cite{Sen}. However, from the viewpoint of experimental realization, it is not always applicable to introduce the time-delayed-feedback into the reaction terms directly. In most of the experimental cases, the time delay acts as an additional feedback and is expressed as additional terms in the reaction-diffusion models \cite{Mikhailov}. This can be found for example in semiconductor devices \cite{Niedernostheide, Beck}, nonlinear optical system consisting of a sodium vapor cell and a single feedback mirror \cite{Lu, Schuttler}, and light-sensitive Belousov-Zhabotinsky reaction \cite{Kheowan, Braune}. So, how about the competition between Turing and Hopf instabilities under time delay in a realistic system? In this paper, we study the role of time delay played on controlling the competition between the Turing and Hopf modes in a model which is originally derived for charge transport in a layered semiconductor system \cite{Niedernostheide, Beck}. We concentrate on the determination of feedback intensities on the oscillation frequency, the intrinsic wave vector, and the intensities of the modes. Results show that time delay with different forms of feedbacks changes the dynamical characteristics near the codimension-two Turing-Hopf bifurcations. Either individual or both feedbacks can realize the transformation between the Turing and Hopf patterns. The corresponding numerical simulations are also done in two dimensions.

\section{MODEL and analysis}

\indent In this work, we study the following two-component reaction-diffusion model:
\begin{eqnarray}
  u_t &=& \frac{v-u}{(v-u)^2+1} - T u + D_{u}\nabla^2 u +F,\\
  v_t &=& \alpha(j_{0}-(v-u))+ D_{v}\nabla^2 v +G,
\end{eqnarray}
here the time delay is applied with the forms:
\begin{eqnarray}
  F=g_{u}(u(t-\tau)-u(t)),\\
  G=g_{v}(v(t-\tau)-v(t)),
\end{eqnarray}
\indent This dimensionless model is derived from the charge transport in a layered semiconductor system \cite{Wacker}. The variable $v$ represents the voltage across the heterostructure and here acts as an activator. The variable $u$ indicates an internal degree of freedom, e.g., the interface charge density, and acts as an inhibitor. The parameters $D_{u}$ and $D_{v}$ describe the diffusion of interfaces charge carriers and the induced inhomogeneity of the voltage drop at the interface, respectively. $T$ is an internal system parameter describing the tunneling rate, and $j_{0}$ is an external current density. The parameter $\alpha$ denotes the ratio of the time scales of $u$ and $v$. The parameters $g_{u}$ and $g_{v}$ are the feedback intensities of variables $u$ and $v$, respectively. $\tau$ is the delayed time. In the analysis, we first consider a short delayed time, i.e. $\tau$ is a small value. So, we can expand the feedback terms Eqs. (3) and (4) as:
\begin{eqnarray}
  u(t-\tau) &=& u(t) -\tau \frac{\partial u(t)}{\partial t},  \\
  v(t-\tau) &=& v(t) -\tau \frac{\partial v(t)}{\partial t}.
\end{eqnarray}
So, we obtain:
\begin{eqnarray}
  (1+\tau g_{u})u_t &=& \frac{v-u}{(v-u)^2+1} - Tu + D_{u}\nabla^2 u,\\
  (1+\tau g_{v})v_t &=& \alpha(j_{0}-(v-u))+ D_{v}\nabla^2 v.
\end{eqnarray}
It can be seen that the terms of time is rescaled and determined by the delayed time and the feedback intensities. In order to study the competition between the Turing and Hopf instabilities under time delay, we first perform the linear stability analysis of Eqs. (7) and (8). The uniform steady states of Eqs. (7) and (8) are determined by the intersection points of the two nullclines:
\begin{eqnarray}
  \frac{v-u}{(v-u)^2+1} - Tu &=& 0,\\
   j_{0}-(v-u) &=& 0.
\end{eqnarray}

\begin{figure}[htbp]
  \begin{center}\includegraphics[width=8cm,height=6cm]{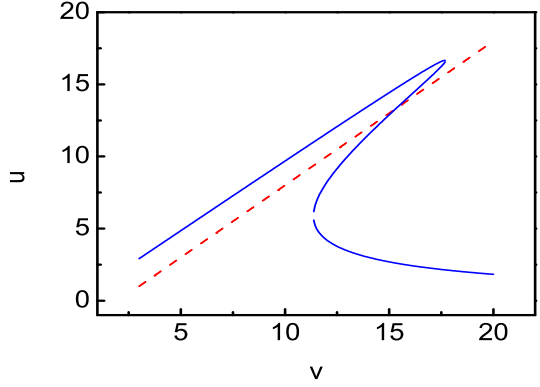}
  \caption{Nullclines of the model system. $D_{u}$$=$$0.1$, $D_{u}$$=$$0.348$, $\alpha$$=$$0.08$, $j_{0}$$=$$2.0$, $T$$=$$0.03$.}\label{1}
  \end{center}
\end{figure}

\indent Obviously, the cubic nullcline of Eq. (9) have a form of "N" as indicated in Fig.1. If the two nullclines intersect at a single point lying on the middle branch of the cubic nullcline, the system exhibits Turing/Hopf instabilities. If they intersect at one point lying on the outer branches of the cubic nullcline, the system owns excitable property. Otherwise they intersect at three points, which corresponds to a bistable system. Here, we only concentrate on the case of Turing/Hopf instabilities. The uniform steady state of the system reads
\begin{eqnarray}
  u_{0} &=& \frac{j_{0}}{T(j_{0}^2+1)},\\
  v_{0} &=& j_{0} + u_{0}.
\end{eqnarray}
Then, we perturb the uniform steady state with small spatiotemporal perturbation ($\delta$$u$, $\delta$$v$) $\sim$ $exp(\lambda t+ikr)$, and obtain the following matrix equation for eigenvalues:
\begin{displaymath}
\mathbf{ }
\left( \begin{array}{cc}
\gamma-$T$ & -\gamma  \\
\alpha & -\alpha
\end{array} \right)
\left( \begin{array}{cc}
\delta u \\
\delta v
\end{array} \right)=0,
\end{displaymath}
where, $\gamma$$=$$\frac{j_{0}^2-1}{(j_{0}^2+1)^2}$. So the characteristic equation reads as
\begin{equation}
A \lambda^2-B\lambda+C = 0,
\end{equation}
where,
\begin{equation}
A = (1+\tau g_{u})(1+\tau g_{v}),
\end{equation}
\begin{equation}
B = (1+\tau g_{u})(-\alpha - D_{v}k^2)+(1 + \tau g_{v})(\gamma-T - D_{u}k^2),
\end{equation}
\begin{equation}
C = (\gamma-T - D_{u}k^2)(-\alpha - D_{v}k^2) + \alpha \gamma.
\end{equation}
The dispersion relations $\lambda(k)$ of the system are defined by the eigenvalues of Eq. (13),
 \begin{equation}
\lambda_{1,2} = \frac{B \pm \sqrt{B^2 -4AC}}{2A}.
\end{equation}
 \indent If the real parts of eigenvalues are positive, the uniform steady state becomes unstable to small perturbations. Here, we focus on the temporal instability (Hopf mode) and spatial instability (Turing mode), which correspond to $k$$=$$0$ and $k$$\neq$$0$, respectively. Figure 2 represents the dispersion relation and shows the effect of time delay on the competition between Turing and Hopf modes. It can be seen that the application of the time delay changes the oscillation frequency of the Hopf instability and the intensities of both the Turing mode and the Hopf mode. Therefore, it affects the results of competition between the Hopf and Turing modes. In the following we will discuss the effect of time delay on the Hopf and Turing instabilities, respectively.

\begin{figure}[htbp]
  \begin{center}\includegraphics[width=8cm,height=6cm]{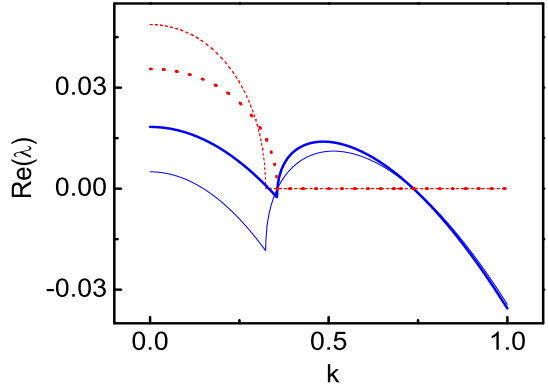}
  \caption{Dispersion relation $\lambda(k)$ with and without time delay. The thick and thin solid lines correspond to the real part of the eigenvalue with and without time delay, respectively. The thick and thin dash lines correspond to the image part of the eigenvalue with and without time, respectively. The feedback parameters are $\tau$$=$$0.1$, $g_{u}$$=$$0.0$, $g_{v}$$=$$5.0$. Other parameters are: $D_{u}$$=$$0.1$, $D_{v}$$=$$0.348$, $\alpha$$=$$0.08$, $j_{0}$$=$$2.0$, $T$$=$$0.03$.}\label{1}
  \end{center}
\end{figure}

\indent For Hopf instability, the most unstable mode occurs at $k$$=$$0$. So the threshold condition of Hopf instability is expressed as
\begin{equation}
\gamma_{H} = T + \frac{\alpha (1 + \tau g_{u})}{1 + \tau g_{v}}.
\end{equation}
  and the corresponding critical frequency of temporal oscillation is
 \begin{equation}
\omega_{0} = \sqrt{\frac{\alpha T}{(1 + \tau g_{u})(1 + \tau g_{v})}}.
\end{equation}

\begin{figure}[htbp]
  \begin{center}\includegraphics[width=8cm,height=12cm]{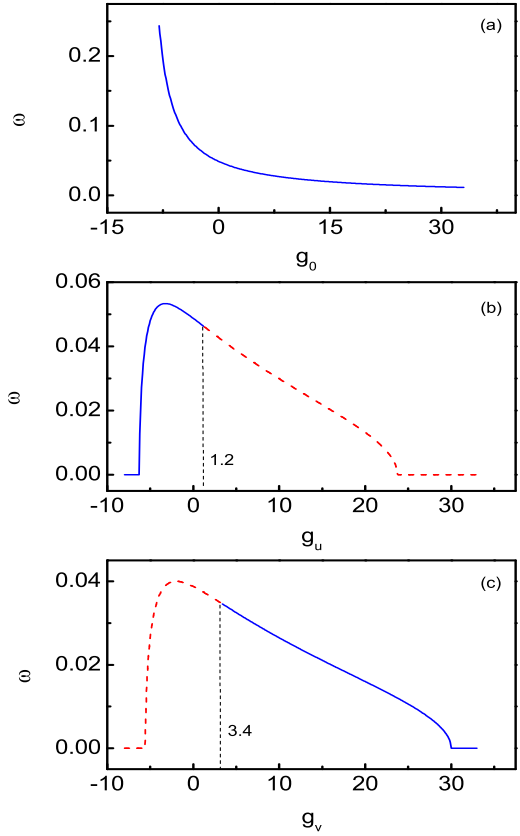}
  \caption{Dependence of the oscillation frequency on the feedback intensities. (a) $g_{u}$$=$$g_{v}$$=$$g_{0}$, (b) $g_{u}$$\neq$$0$, $g_{v}$$=$$0$. (c) $g_{u}$$=$$0$, $g_{v}$$\neq$$0$. The solid and dash lines correspond to the cases of Re($\lambda$)$>$$0$ and Re($\lambda$)$<$$0$, respectively. $D_{u}$$=$$0.1$, $D_{v}$$=$$0.348$, $\alpha$$=$$0.08$, $j_{0}$$=$$2.0$, $T$$=$$0.03$, $\tau$$=$$0.1$.}\label{1}
  \end{center}
\end{figure}

  \indent From Eq. (18) it can be seen that the time delay changes the critical value of Hopf bifurcation when the feedback $g_{u}$ and $g_{v}$ are applied with different intensities. This means that time delay with different feedback intensities can introduce the Hopf instability to system, which is similar to the cases in refs. [13, 14]. However, application of the feedback with identical intensities does not alter the point of Hopf bifurcation, and it only changes the oscillatory frequency as indicated by Eq. (19). In Eq. (19), ($1 + \tau g_{u}$)($1 + \tau g_{v}$) must be larger than zero in order to make $w_{0}$ a real value when applying the time-delayed-feedback. Figure 3 (a) shows the dependence of oscillatory frequency $w$ on the feedback intensity when applying identical intensities of feedback, such that $g_{u}$$=$$g_{v}$$=$$g_{0}$. It shows that $w$ decreases as the feedback intensity increases. Figure 3 (b) and (c) represent the dependence of oscillation frequency on the parameter $g_{u}$ and $g_{v}$, respectively. The solid lines indicate that the real part of Hopf mode is positive. It can be seen that the Hopf bifurcation occurs within a range of feedback intensity $g_{u}$ and $g_{v}$. If an individual feedback $g_{u}$ is applied, the oscillation frequency first increases and then decreases with $g_{u}$. Within the two critical points $g_{u}$=[$-6.2$, $1.2$], the Hopf bifurcation occurs. If an individual feedback $g_{v}$ is applied, the oscillation frequency decreases with $g_{v}$. There are two critical points $g_{v}$=[$3.4$, $29.9$] within which the Hopf bifurcation occurs. Figure 3 (b) and (c) further confirm that the time delay can induce Hopf instability if $g_{u}$$\neq$$g_{v}$.

\begin{figure}[htbp]
  \begin{center}\includegraphics[width=8cm,height=6cm]{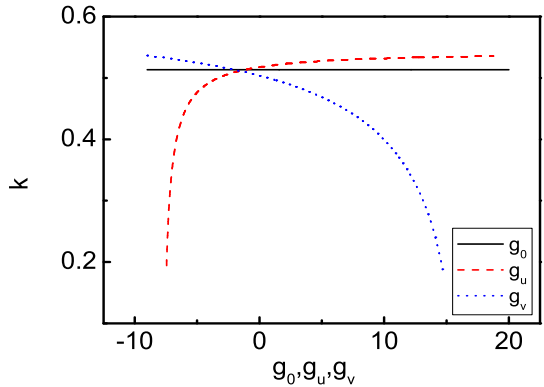}
  \caption{Dependence of the intrinsic wave vector on the feedback intensities. Solid line: $g_{u}$$=$$g_{v}$$=$$g_{0}$, dash line: $g_{u}$$\neq$$0$, $g_{v}$$=$$0$, dot line: $g_{u}$$=$$0$, $g_{v}$$\neq$$0$. $D_{u}$$=$$0.1$, $D_{v}$$=$$0.348$, $\alpha$$=$$0.08$, $j_{0}$$=$$2.0$, $T$$=$$0.03$, $\tau$$=$$0.1$.}\label{1}
  \end{center}
\end{figure}

\indent For Turing instability, we first focus on the critical wave vector $k_{0}$ and the critical value for Turing bifurcation. The bifurcation threshold for Turing instability is $\gamma_{T}$$=$${[{T}^{1/2}+{(\frac{D_{u}\alpha}{D_{v}})}^{1/2}}]^{1/2}$. At the Turing bifurcation point the critical wave vector $k_{0}$ is not affected by the time delay, and it reads $k_{0}^{2}$=${( \frac{\alpha T}{D_{u} D_{u}})}^{1/2}$. So the time delay does not change the critical wave vector and the critical value for Turing bifurcation. This is due to the fact that the time delay is equivalent to rescaling time as shown in Eqs. (7) and (8). Beyond the Turing bifurcation, the time delay alters the intrinsic wave vector and the intensity of Turing mode, i.e. the maximum real part of the eigenvalues as shown in Fig. 2. The dependence of the intrinsic wave vector on the feedback intensity is shown in Fig. 4. If an individual feedback $g_{u}$ is applied, such that $g_{u}$$\neq$$0$, $g_{v}$$=$$0$, the wave vector increases monotonously with feedback intensity $g_{u}$ as indicated by the dash line in Fig. 4. If an individual feedback $g_{v}$ is applied, such that $g_{u}$$=$$0$, $g_{v}$$\neq$$0$, the wave vector decreases monotonously with feedback intensity $g_{v}$ as indicated by the dot line in Fig. 4. However, if the competition between the feedback $g_{u}$ and $g_{v}$ reaches balance $g_{u}$$=$$g_{v}$$=$$g_{0}$, the wave vector remains constant while changing the feedback intensities, as shown by the solid line in Fig. 4. The three lines intersect at one point at the coordinate ($g_{0}$, $k$)=($-1.6$, $0.51$).

\begin{figure}[htbp]
  \begin{center}\includegraphics[width=8cm,height=12cm]{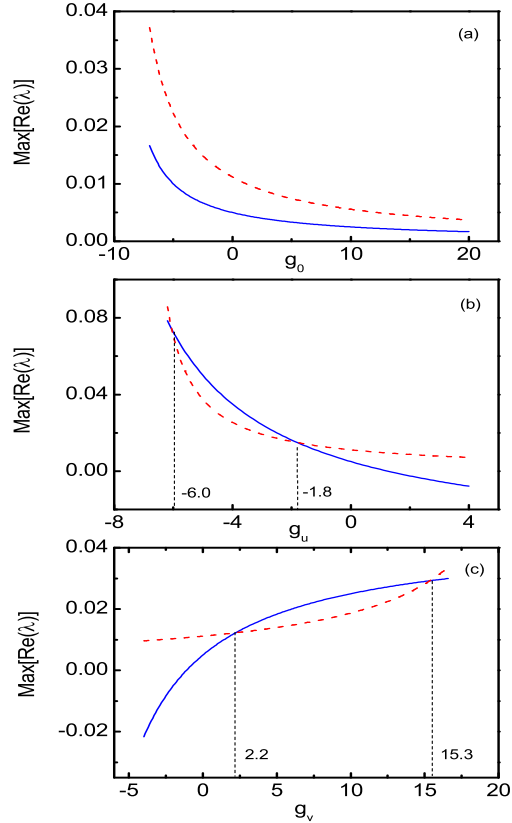}
  \caption{Dependence of the mode intensities on the feedback intensities. (a) $g_{u}$$=$$g_{v}$$=$$g_{0}$, (b) $g_{u}$$\neq$$0$, $g_{v}$$=$$0$. (c) $g_{u}$$=$$0$, $g_{v}$$\neq$$0$. The solid and dash lines correspond to the Hopf and Turing modes, respectively. $D_{u}$$=$$0.1$, $D_{v}$$=$$0.348$, $\alpha$$=$$0.08$, $j_{0}$$=$$2.0$, $T$$=$$0.03$, $\tau$$=$$0.1$.}\label{1}
  \end{center}
\end{figure}

\indent Near the codimension-two Turing-Hopf bifurcations, the Turing mode and the Hopf mode compete. The competetition depends on the intensities of the two modes. Figure 5 shows the dependence of the most unstable modes (both Turing and Hopf) on the feedback intensities. Here, we use the set of parameters in Fig. 1. With this set the Turing mode is stronger than the Hopf mode in the absence of time delay. When the feedbacks are applied with identical intensities $g_{u}$$=$$g_{v}$$=$$g_{0}$, the two maximum real part of the eigenvalues which correspond to the Hopf and Turing modes decreases with feedback intensities as shown in Fig. 5 (a). The Turing mode remains stronger than the Hopf mode. If an individual feedback $g_{u}$ is applied, such that $g_{u}$$\neq$$0$, $g_{v}$$=$$0$, the intensities of the two modes decrease with $g_{u}$ as shown in Fig. 5 (b). We want to point out that in this case, the Turing mode is able to become weaker than the Hopf mode within the range $-6.0$$<$$g_{u}$$<$$-1.8$, which differs greatly from Fig. 5 (a). This means that the Turing pattern could transit into Hopf pattern if feedback is applied with appropriate intensity. If an individual feedback $g_{v}$ is applied, such that $g_{u}$$=$$0$, $g_{v}$$\neq$$0$, the intensities of the two modes increase with $g_{v}$ as shown in Fig. 5 (c). There exists a parameter range $2.2$$<$$g_{v}$$<15.3$, within which the Hopf mode is stronger than the Turing mode.

\indent So it can be seen that by using time-delay-feedback one can realize the transition between the spatial Turing pattern and the temporal Hopf pattern, which depends on how the intensities of feedbacks are applied. It is practicable to control the self-organized patterns by applying individual or both of the feedbacks. As a result of competition between the Turing and Hopf modes controlled by time delay as shown in Fig. 5, an original stronger Turing mode becomes weaker and then stronger than Hopf mode with increasing individual feedback intensity continuously.

\section{Two-dimensional numerical simulation}
\indent In view of the aforesaid analysis we carry out numerical simulation by using a generalized Peaceman-Rachford ADI scheme on a square grid of $100$$\times$$100$ space units. The time step is $dt$$=$$0.02$ time unit and the space step is $dx$$=$$dy$$=$$1.0$ space unit. The boundary conditions are taken to be no-flux.

\begin{figure}[htbp]
  \begin{center}\includegraphics[width=8.5cm,height=8cm]{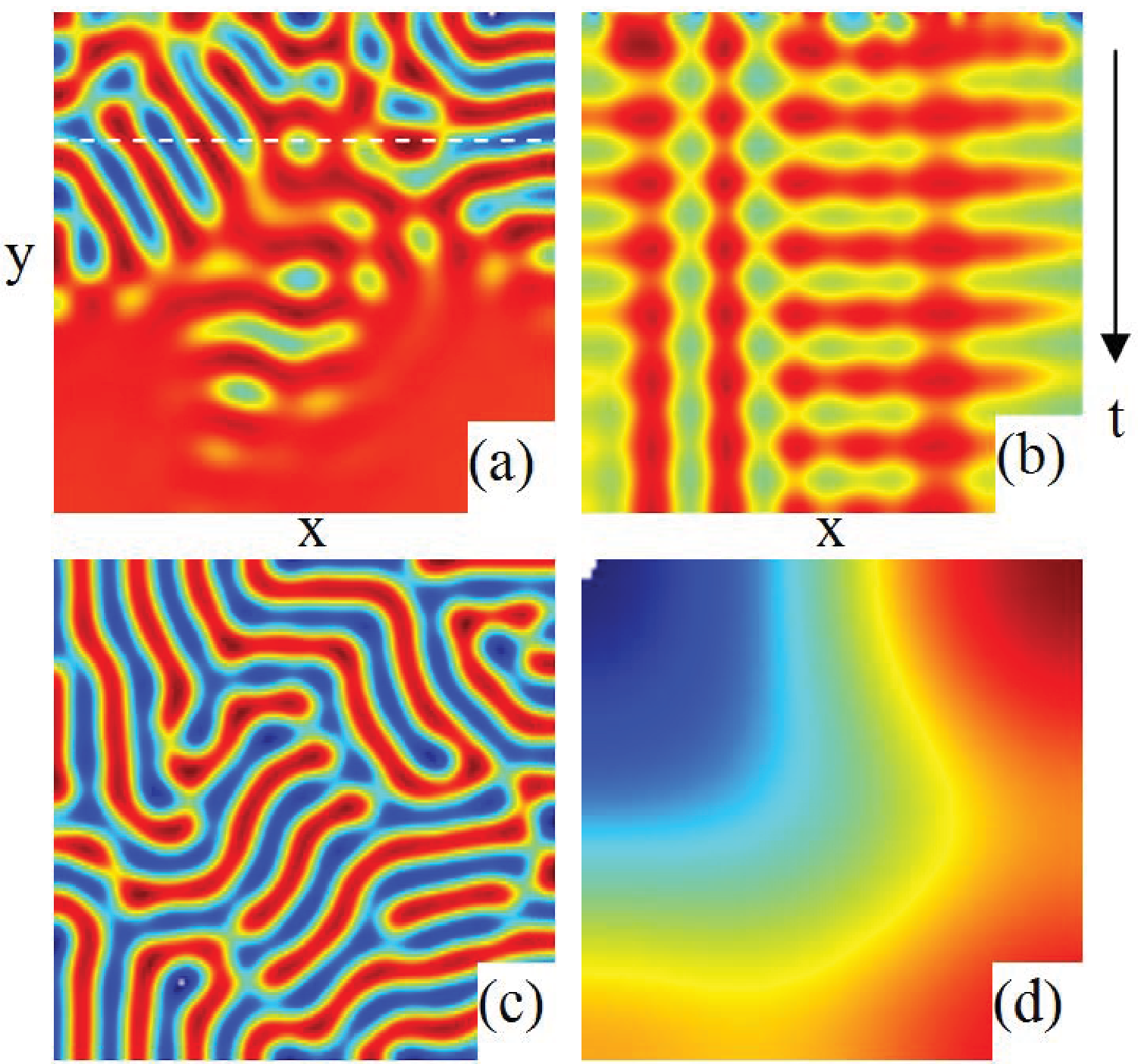}
  \caption{Two-dimensions numerical simulation. (a)coexistence of the Turing stripe and the Hopf oscillation in the absence of time delay. (b) time evolution of variable $u$ along the white dash line indicated in (a). (c) stripe pattern under time delay, $\tau$$=$$0.1$, $g_{u}$$=$$5.0$, $g_{v}$$=$$0.0$. (d) Hopf oscillation under time delay, $\tau$$=$$0.1$, $g_{u}$$=$$0.0$, $g_{v}$$=$$5.0$. The other parameters are $D_{u}$$=$$0.1$, $D_{v}$$=$$0.348$, $\alpha$$=$$0.08$, $j_{0}$$=$$2.0$, $T$$=$$0.03$.}\label{1}
  \end{center}
\end{figure}

\indent In order to keep the validation of the Eqs. (7) and (8), we first use a small delayed time $\tau$$=$$0.1$ in Fig. 6. However, extensive numerical simulations have shown that when applying feedback with long delayed time and weak intensity, e.g. $\tau$$=$$10.0$, $g_{u}$$=$$0.1$, the analysis results still remain correct qualitatively because the delayed time and the feedback intensity are incorporated into the coefficient of the left-hand terms of the Eqs. (7) and (8).

\indent We concentrate on the competition between the Turing pattern and the Hopf pattern by applying time delay. Fig. 6 (a) first shows a coexistence of Turing stripe and Hopf oscillation near the codimension-two Turing-Hopf bifurcations in the absent of time delay. The time evolution along the marked white dash line in Fig. 6 (a) confirms the stability of the coexistence state as shown in Fig. 6 (b). If an individual feedback $g_{u}$$=$$5.0$ is applied, as predicted previously in Fig. 5 (b), the Turing mode becomes stronger than the Hopf mode. So we obtain pure Turing stripe as shown in Fig. 6 (c). However, if we apply the feedback $g_{u}$$=$$-5.0$, the Hopf mode plays a leading role. So we observe oscillatory pattern as shown in Fig. 6 (d).

\indent If an individual feedback $g_{v}$$=$$5.0$ is applied, the original Fig. 6 (a) evolves into Hopf oscillation as Fig. 6 (d) which shows the dominance of Hopf mode. However, as indicated in Fig. 5 (c), if we use $g_{v}$$=$$-2.5$, the Turing mode becomes dominant. We obtain stripe pattern like Fig. 6 (c). So the application of the time delay can realize the transformation between the Turing and Hopf instabilities resulting from the competition between the Turing and Hopf modes.

\section{CONCLUSIONS}
\indent We have studied the competition of the spatial and temporal instabilities under time delay near codimension-two Turing-Hopf bifurcations in a two-components reaction-diffusion equation. The time-delayed-feedback is determined by the delayed time and the feedback intensities. Based on the linear stability analysis, we have investigated the effect of time delay on the Turing and Hopf instabilities. Results have shown that appropriate time delay can introduce a Hopf bifurcation when the intensities of feedback are different. It also affects the oscillation frequency of Hopf mode. However, the time delay does not change the bifurcation threshold and the corresponding critical wave vector for Turing instability. Beyond the Turing bifurcation, the time delay begins to work on controlling the intrinsic wave vector and the intensity of Turing mode. The wave vector increases (decreases) with the feedback intensity when individual feedback $g_{u}$ (or $g_{v}$) is applied. If the feedback intensities $g_{u}$ and $g_{v}$ are equal, the time delay does not change the intrinsic wave vector. Near the codimension-two Turing-Hopf bifurcations the competition between the Turing and Hopf modes results in the transition between the Turing and Hopf patterns. The intensities of both Turing and Hopf modes decrease (increase) with the intensity of feedback $g_{u}$ ($g_{v}$). Analysis shows that individual or both feedbacks can realize the control of the transformation between Turing and Hopf patterns. As a result of competition between the Turing and Hopf modes controlled by time delay, initially stronger Turing mode becomes weaker and then stronger than Hopf mode with increasing individual feedback intensity continuously. We have carried out two dimensional numerical simulation which has validated the analytical results. Extensive numerical simulations have also shown that when applying feedback with long delayed time but weak intensity, the results are still correct qualitatively. We hope that our results could be instructive to experiments, such as the charge transport in semiconductor devices and the light-sensitive Belousov-Zhabotinsky chemical reactions.

\section{ACKNOWLEDGMENTS}
\indent  This work is supported by the Fundamental Research Funds for the Central Universities (No. 09ML56) and the Foundation for Young Teachers of the North China Electric Power University, China (No. 200611029).

\end{document}